\documentclass[twocolumn]{aastex701}

\usepackage{amsmath}
\graphicspath{{./}{figures/}}
\usepackage{threeparttable}
\usepackage{booktabs}

\shorttitle{Source Approximations in Surface Flux Transport Simulations}
\shortauthors{Yukun Luo et al.}

\begin{document}

\title{Effects of Various Bipolar Approximations of Active Regions on Solar Surface Magnetic Field Simulations}

\author{Yukun Luo}
\affiliation{School of Space and Earth Sciences, Beihang University, Beijing, People’s Republic of China}
\affiliation{Key Laboratory of Space Environment Monitoring and Information Processing of MIIT, Beijing, People’s Republic of China}
\email{luoyukun@buaa.edu.cn}

\author{Jie Jiang}
\affiliation{School of Space and Earth Sciences, Beihang University, Beijing, People’s Republic of China}
\affiliation{Key Laboratory of Space Environment Monitoring and Information Processing of MIIT, Beijing, People’s Republic of China}
\email[show]{jiejiang@buaa.edu.cn}

\author{Ruihui Wang}
\affiliation{School of Space and Earth Sciences, Beihang University, Beijing, People’s Republic of China}
\affiliation{Key Laboratory of Space Environment Monitoring and Information Processing of MIIT, Beijing, People’s Republic of China}
\email{wangruihui@buaa.edu.cn}


\begin{abstract}
The evolution of solar surface magnetic fields is essential for understanding solar activity and the underlying dynamic process. The surface flux transport (SFT) model is a widely used and effective tool for simulating this evolution. Active regions are incorporated as magnetic flux sources of the SFT model, but their configurations are usually simplified as symmetric or asymmetric bipolar magnetic regions (BMRs). Here, we aim to quantitatively and systematically assess how such flux source approximations affect SFT results and explore improved approximation methods using our recently developed SFT code. By comparing simulations that incorporate realistic active region configurations from solar cycle 23 through the ongoing cycle 25, we show that approximating active regions as symmetric BMRs leads to a systematic overestimation of the axial dipole strength at solar minimum. This result is independently confirmed using an algebraic quantification that evaluates the axial dipole contribution of individual active regions. The overestimation can be partially reduced by monotonically decreasing the size of the approximated BMRs, but it cannot be fully eliminated. When active regions are instead represented by morphologically asymmetric BMRs, the simulated axial dipole strength exhibits a strong and nearly linear negative dependence on the size ratio between the following and leading polarities. Based on these results, we propose a combination of BMR size and polarity size ratio that yields an axial dipole evolution comparable to that obtained with fully incorporated realistic active region configurations. This study provides a new quantitative constraint for improving future simulations with approximated BMRs.
\end{abstract}


\section{Introduction}\label{sec:intro} 
The solar surface magnetic field is an important component of the solar magnetic field, and its evolution plays a crucial role in the Babcock-Leighton dynamo process \citep{Babcock1961, Zhang2022} as well as in modeling heliospheric magnetic fields \citep{Wang2011, Wang2014}. The surface flux transport (SFT) model initially introduced by \cite{Leighton1964} provides a powerful tool for investigating the evolution of surface fields. Over decades of development, previous studies have demonstrated that the SFT model can successfully reproduce the observed large-scale surface magnetic field and is widely applied to solar cycle prediction, as reviewed by \cite{Sheeley2005, Jiang2014, Wang2017, Yeates2023, Petrovay2020pre}.

The SFT model incorporates the radial magnetic flux that emerges at the solar surface as the source term driving subsequent evolution. This flux is then transported by horizontal flows, including differential rotation, meridional flow, and supergranular diffusion, to reverse or build the solar polar field. As transport parameters are not well constrained by observations, several studies investigated the influence of transport parameters \citep{Devore1984, Wang1989, Baumann2004} and constrained their optimal ranges through the comparisons of simulated and observed features, such as the polar fields \citep{Lemerle2015, Petrovay2019} and magnetic power spectra \citep{Luo2025}. Besides the transport parameters, the source term is also crucial in determining SFT outcomes \citep{Jiang2023, Yeates2023}.

\cite{Jiang2019} propose the importance of incorporating active regions (ARs) with realistic configurations as the source term for the first time. \cite{Yeates2015, Wang2025} incorporate the ARs identified from synoptic magnetograms and successfully reproduce the observed poleward surges. \cite{Yeates2025} use the historical database reconstructed by Ca \uppercase\expandafter{\romannumeral2} K synoptic maps to successfully reproduce the surface magnetic field evolution across several solar cycles without the additional radial diffusion term suggested by \cite{Schrijver2002, Baumann2006}. However, identifying ARs from magnetograms is difficult and time-consuming, and the limited temporal coverage of observations restricts their application. For instance, in the work of reconstructing historical magnetograms over several centuries \citep{Jiang2011a, Hofer2024}, the sunspot number record is the only long-term proxy available. Hence, simplified forms of the source term are necessary to be adopted when realistic AR data are unavailable.

Most previous studies simplify ARs as morphological symmetry bipolar magnetic regions (BMRs), such as \cite{vanBallegooijen1998}. The configuration of the generated BMR could be determined by its tilt angle, location, area, and so on, based on observed statistical properties \citep{Jiang2011a, Jiang2018, Yeates2020}. Although incorporating such BMRs can reproduce many observed features \citep{Cameron2010, Jiang2011}, a scaling factor of 0.7 for the tilt angle is required in their simulations. Additionally, \cite{Yeates2020} find that approximated BMRs tend to overestimate the simulated axial dipole strength, a widely used proxy for the polar field, relative to simulations using realistic AR configurations under identical transport parameters. This conclusion is independently supported by \cite{Wang2021, Wang2024}, who evaluate the axial dipole strength through the algebraic method. Furthermore, \cite{Jiang2019} demonstrate that the evolution of complex ARs changes and even provides the opposite contribution to the axial dipole strength after the BMR approximation, highlighting the sensitivity of the SFT simulation to the source configuration.

Observations indicate that the following polarity is generally more dispersed than the leading polarity \citep{Fisher2000, Tlatov2014, Fan2021}. This morphological feature is also reproduced in the numerical simulations \citep{Fan1993, Caligari1995, Rempel2014}. The importance of this asymmetry is emphasized by \cite{Iijima2019}, who show that asymmetric sources in a one-dimensional SFT model yield results more consistent with observations than symmetric ones. They further constrain source asymmetry indirectly using observational statistics from \cite{Tlatov2014}. Based on their investigation, \cite{Pal2025b} reconstruct the century-long solar magnetic field by incorporating asymmetry sources. Meanwhile, \cite{Wang2021} reveal the weakness of the traditional BMR-based algebraic method when considering morphological asymmetry.

These studies collectively illustrate that the configuration of the flux source plays a central role in SFT simulations. However, a systematic comparison between various flux source approximations and realistic ARs is still lacking. Here, we perform a more comprehensive assessment of how approximating realistic ARs as BMRs with varying sizes and degrees of asymmetry affects the evolution of the surface polar field over solar cycles 23-25. This analysis is based on the continuous and homogeneous identified realistic AR database obtained by \cite{Wang2023, Wang2024} and aims to improve the existing BMR approximation methods. We quantify these effects using the axial dipole strength as the quantitative index, employing our SFT code developed by \cite{Luo2025}. This code, based on spherical harmonic decomposition, enables high-accuracy and high-efficiency simulations and facilitates investigations of the multiscale features of surface magnetic fields like \cite{Luo2023, Luo2024}.

The paper is organized as follows. In Section \ref{sec:model}, we describe our SFT code, parameter sets, method for incorporating source terms, and BMR approximation methods. Section \ref{sec:results} presents and compares the simulation results obtained with different flux source configurations. In Section \ref{sec:conclusion}, we summarize our results and discuss their implications.

\section{Methods}\label{sec:model}

\subsection{Surface Flux Transport Model} \label{subsec:SFT}  
The evolution of surface radial magnetic field $B(\theta,\phi,t)$ is described by the two-dimensional SFT equation:
\begin{equation}\label{eq1}
\frac{\partial B}{\partial t}+\nabla\cdot(U_sB) = \eta\nabla_s^2 B+S(\theta, \phi, t),	
\end{equation}	
where $\theta, \phi, t$ are colatitude, longitude, and time, respectively. The parameter $U_s$ denotes the large-scale advection surface flow field, consisting of the differential rotation and the meridional flow $v(\theta)$. We adopt the differential rotation profile from \cite{Snodgrass1983}, which is widely used in previous work \citep{Baumann2006, Jiang2014}. 

The meridional flow profile is modified from the formulation of \cite{Lemerle2015} for better simulation of polar fields and is expressed as
\begin{equation}\label{eq2}
    v(\theta) = -v_0\ \mathrm{erf}^8(2.25\ \sin(\theta))\ \mathrm{erf}(3.5\ \cos(\theta)),
\end{equation}
where $v_0$ controls the overall flow amplitude, including peak flow speed and divergence at the equator $\Delta v$. In this study, we adopt $v_0 \approx 13\ m/s$, which is within the range of common meridional flow measurements. The value of $\Delta v$ is approximately 1 m$\cdot$s$^{-1} \cdot$deg$^{-1}$, which is close to the results measured by Doppler shift measurements \citep{Ulrich2010} and helioseismology \citep{Zhao2014, Liang2018}, as listed in Table 1 of \cite{Jiang2023}. This profile yields a relatively slight decrease in flow speed at mid-latitudes. Additional details on constraining this meridional flow formulation will be presented in a forthcoming paper. 

The effect of supergranulation is represented by a turbulent diffusion term \citep{Leighton1964}, with $\eta$ denoting the corresponding diffusivity. The term $S(\theta, \phi, t)$ describes the emerged radial magnetic flux due to the internal process. The different methods for incorporating flux sources employed in this work are introduced in the following subsections.

To perform the simulations, we use the recently developed SFT code \citep{Luo2025}, which solves Equation (\ref{eq1}) using a spectral method based on spherical harmonic decomposition and a fourth-order Runge–Kutta scheme. In this paper, the initial synoptic map and source magnetograms are represented on a 360 $\times$ 180 grid with uniform spacing in longitude and latitude. We adopt a maximum spherical harmonic degree of $l_{max} = 60$, following the suggestions of \cite{Luo2025}, and the time step $\Delta t =$ 1 day. 

\subsection{Incorporating Realistic AR Configurations as Flux Sources} \label{subsec:real}

At first, we incorporate ARs using their realistic configurations as flux sources. The target ARs are taken from the Active Region database for Influence on Solar cycle Evolution (ARISE; \cite{Wang2023, Wang2024}). This database covers Carrington Rotation (CR) 1909 to the latest CR and detects ARs from Solar Dynamics Observatory (SDO)/Helioseismic and Magnetic Imager (HMI) \citep{Scherrer2012, Scherrer2012b} and Solar and Heliospheric Observatory (SOHO)/Michelson Doppler Imager (MDI) synoptic maps \citep{Scherrer1995}. The database and associated codes are publicly available on GitHub \footnote{\texttt{AR database:} \url{https://github.com/Wang-Ruihui/A-live-homogeneous-database-of-solar-active-regions}.} and version 4.0 is archived in Zenodo (\dataset[10.5281/zenodo.15076075]{https://doi.org/10.5281/zenodo.15076075}; \citealt{Database_WangRH_2025}). In this study, we incorporate ARs from CR 1911 to CR 2297 (1996 July–2025 April), covering all of cycles 23 and 24 and part of cycle 25.

All magnetograms containing ARs from ARISE are smoothed, and their spatial resolution is reduced to 360 $\times$ 180 prior to their incorporation as source terms. According to the tests of \cite{Luo2025}, adopting the higher spatial resolution of initial magnetogram and flux source maps or a higher maximum spherical harmonic degree would not affect the large-scale simulated results, while increasing the computational cost of spherical-harmonic decomposition. In other words, the simulation results during $l_{max}=60$ and the resolution of 360 $\times$ 180 are convergent and reliable with high efficiency.

Following \cite{Luo2025}, we use the same framework to incorporate the newly emerged magnetic region into the simulated magnetograms. For details, the same source-term incorporation cadence is adopted, whereby all ARs emerging within a given CR are incorporated on the final day of that CR. This cadence could reduce the discrepancies between simulated and observed synoptic maps. A notable difference is that the newly emerged ARs are added to the simulated magnetograms, rather than replacing the corresponding regions. The same method and cadence for incorporating source terms are also adopted for the BMR sources presented in Section \ref{subsec:approximation}.

Because the simulated axial dipole strength is sensitive to the net surface magnetic flux, it is necessary to balance the flux of each incorporated AR. Here, we use the flux balance method as described in \cite{Yeates2015}. Separate scaling factors for the positive and negative polarity pixels, $S_P$ and $S_N$, are calculated by
\begin{equation}\label{eq3}
    S_P = \frac{\phi_{uns}}{2 \phi_P}, S_N=-\frac{\phi_{uns}}{2\phi_N},
\end{equation}
where $\phi_P$, $\phi_N$, and $\phi_{uns}$ are positive flux, negative flux, and total unsigned magnetic flux of the incorporated AR, respectively. This balance method keeps the total unsigned magnetic flux constant to reduce the potential artifacts.

\subsection{BMR Approximations of ARs}\label{subsec:approximation}
\subsubsection{Model of Symmetric BMRs}\label{subsubsec:sym}

\begin{figure}[!h]
	\centering
	\includegraphics[width=1.0\linewidth]{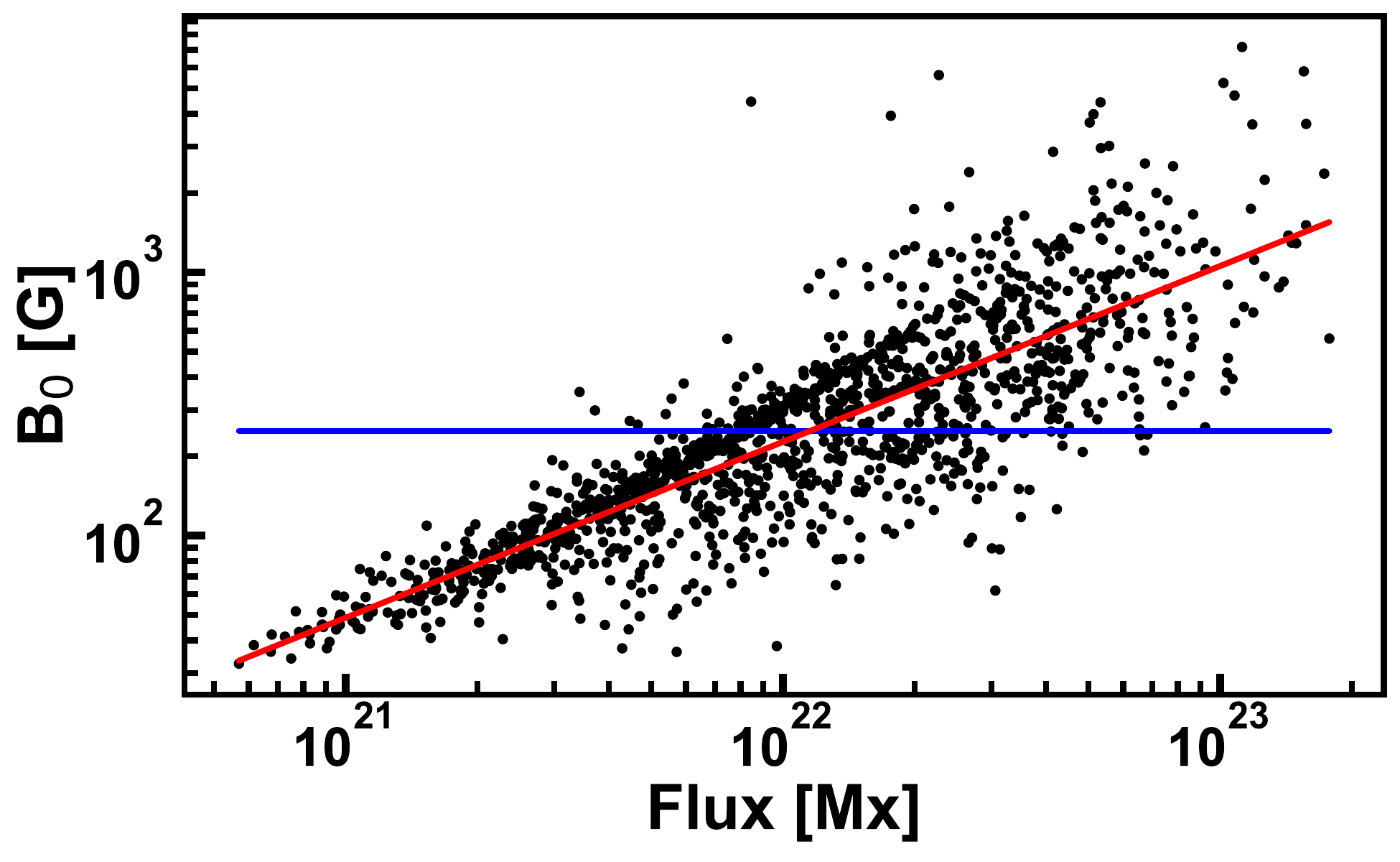}
	\caption{Relation between the scaling factor $B_0$ and the unsigned magnetic flux $\Phi$. The red line is the best-fit relation (Equation~\ref{eq8}), while the blue line indicates the fixed value $B_0 = 250,\mathrm{G}$.}
	\label{fig1}
\end{figure}

In this study, a symmetric BMR is constructed using the given polarity locations $(\theta_\pm, \phi_\pm)$ and the unsigned magnetic flux $\Phi$ as inputs. The source term $S(\theta, \phi, t)$ in Equation (\ref{eq1}) at time $t$ is defined as
\begin{equation}\label{eq4}
    S(\theta, \phi, t) = B^+(\theta, \phi, t) - B^-(\theta, \phi, t),
\end{equation}
where $B^\pm(\theta, \phi, t)$ denote the distributions of newly emerged positive and negative polarities of the BMR, respectively. Following \cite{vanBallegooijen1998, Baumann2006}, each polarity is assumed to have a Gaussian spatial profile,
\begin{equation}\label{eq5}
    B^\pm(\theta, \phi, t) = B_0exp\{-\frac{2[1-cos\beta_\pm(\theta, \phi, t)]}{\delta^2}\},
\end{equation}
where
\begin{equation}\label{eq6}
    cos\beta_\pm=cos\theta_\pm cos\theta+sin\theta_\pm sin\theta cos(\phi-\phi_\pm).
\end{equation}
Here, $\beta_\pm$ denote the heliocentric angular distance between the polarity locations $(\theta_\pm, \phi_\pm)$ and any location ($\theta,\phi$) of the computational grid. The heliocentric angle between $(\theta_\pm, \phi_\pm)$ is the polarity separation, $\Delta \beta$. These definitions yield a BMR composed of two circular polarities.

The parameter $\delta$ represents the angular width (i.e., the radius) of each polarity and directly determines the area of the approximated BMR. It is assumed to scale with the polarity separation as
\begin{equation}\label{eqdelta}
    \delta=k\Delta \beta.
\end{equation}
The proportionality coefficient $k$ is a free input parameter and taken to be 0.4 by \cite{vanBallegooijen1998}, and other subsequent studies usually follow their choice \citep[e.g.,][]{Baumann2004}. The influence of different coefficients would be discussed in Section \ref{subsec:symmetry}. Due to the spatial resolution of the incorporated magnetograms and the truncation spherical harmonic degree, we apply a minimum width of $\delta_{min} = 2^\circ$. If one calculated $\delta$ is smaller than this threshold, its value is replaced by $\delta_{min}$. This threshold results in a polarity with the size of $4^\circ$ that can be adequately resolved with the adopted truncation $l_{max}=60$.

The scaling factor $B_0$ controls the total magnetic flux of the generated BMR. Previous studies typically assume a constant $B_0$, motivated by the positive correlation between area and unsigned magnetic flux \citep{vanBallegooijen1998}. In contrast, we determine $B_0$ individually for each flux source by matching the unsigned magnetic flux of the approximated BMR to input $\Phi$, following the method of \cite{Yeates2020}. As shown in Figure \ref{fig1}, the resulting $B_0$ values exhibit a strong correlation with the unsigned magnetic flux $\Phi$ from realistic ARs, which can be described by the empirical relation
\begin{equation}\label{eq8}
    B_0={10^{-12.34}}\Phi^{0.668}.
\end{equation}
Compared with the commonly used fixed value $B_0=250 \ G$ adopted by \cite{vanBallegooijen1998} and \cite{Baumann2006} (blue line in Figure \ref{fig1}), Equation (\ref{eq8}) yields lower $B_0$ for weak ARs and higher values for strong ARs. This indicates that assuming a constant $B_0$ would tend to overestimate the flux of small ARs while underestimating that of large ones. 

\subsubsection{Model of Asymmetric BMRs}\label{subsubsec:asym}

For the asymmetric BMR approximation, asymmetry is introduced by adjusting the size ratio $f$ between the following and leading polarities, while all other procedures remain the same as in the symmetric BMR approximation described in the last subsection. Following recent work incorporating asymmetry BMRs \citep{Iijima2019, Wang2021}, we adopt the same definition of the size ratio $f$,
\begin{equation}\label{eq7}
    f=(\frac{\delta_F}{\delta_L})^2,
\end{equation}
where $\delta_F$ and $\delta_L$ are the spatial sizes of the following and leading polarities, respectively. The size of the leading polarity is set equal to that in the symmetric BMR case (Section \ref{subsubsec:sym}), i.e., $\delta_L=k\Delta\beta$. Accordingly, the size of the following polarity is then given by $\delta_F = k\Delta\beta\times f^{0.5}$. 

As another free input parameter, $f>1$ results in the following polarity of the approximated BMR being larger than the leading polarity, consistent with observations. Larger values of $f$ correspond to stronger morphological asymmetry. In particular, the case $f=1$ reduces to the symmetric configuration discussed in Section \ref{subsubsec:sym}.


\subsubsection{Approximating observed ARs as BMRs}\label{subsubsec:ARtoBMR}

\begin{figure*}[!ht]
	\centering
	\includegraphics[width=1.0\linewidth]{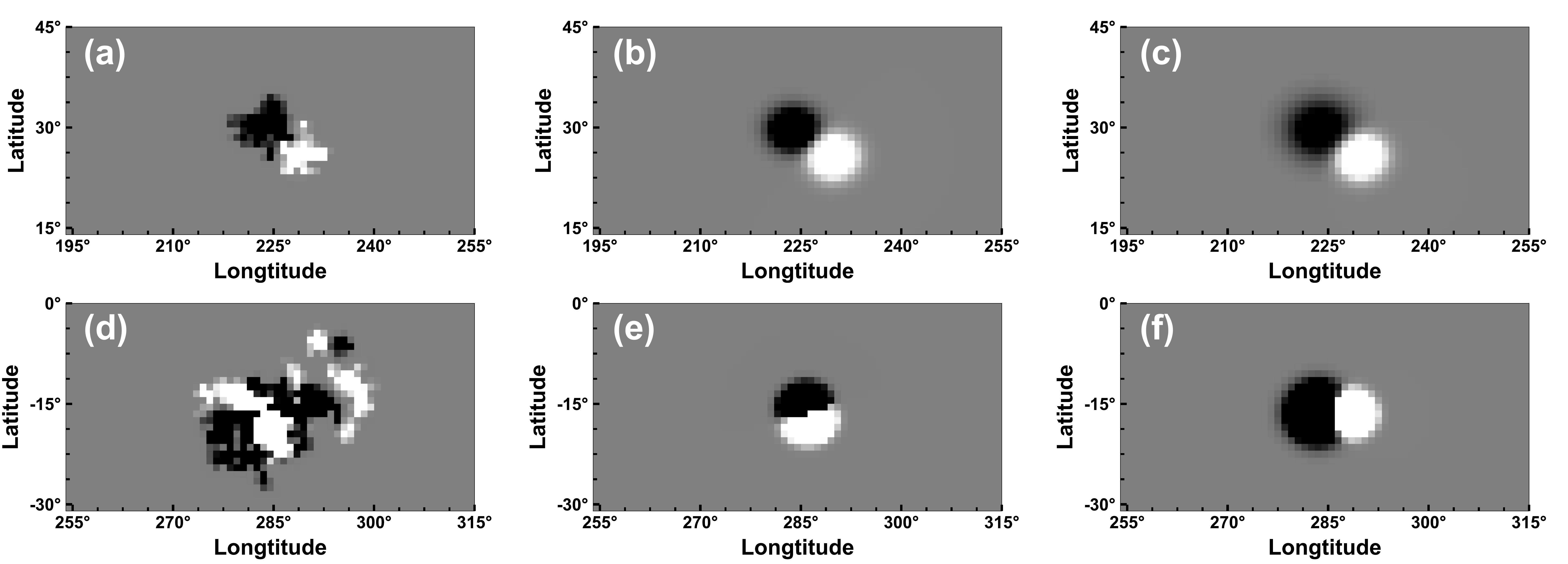}
	\caption{Three incorporated flux source configurations with realistic ARs, symmetry approximated BMR, and asymmetry approximated BMR. NOAA 08086 (top) and NOAA 10486 (bottom) are taken as examples. All magnetograms are saturated at 100 G. (a) Identified realistic configuration of NOAA 08086 from ARISE. (b) Symmetric BMR approximation of NOAA 08086 with $k=0.4$. (c) Asymmetric BMR approximation of NOAA 08086 with $f=2$ and $k=0.4$. (d)-(e) Same as (a)-(c), but for NOAA 10486.}
	\label{fig2}
\end{figure*}

In this subsection, we adopt two representative ARs, NOAA 08086 and NOAA 10486, to illustrate how to approximate realistic ARs as symmetric or asymmetric BMRs with input parameters ($\theta_\pm,\phi_\pm$), $\Phi$, and two free parameters ($k$ and $f$). NOAA 08086 is a typical simple AR, whereas NOAA 10486 exhibits a highly complex configuration. The locations of their polarities ($\theta_\pm,\phi_\pm$) are calculated as the flux-weighted centers. Another input parameter, $\Phi$, is taken directly as the unsigned magnetic flux of each AR. The approximation results are shown in Figure \ref{fig2}. The left panels show the realistic AR configurations, the middle panels display their symmetric BMR approximations with $k=0.4$, and the right panels show the corresponding asymmetric approximated BMRs with $f=2$ and $k=0.4$. 

For NOAA 08086, the symmetric BMR approximation produces a compact and nearly circular structure (Figure \ref{fig2} (b)) with an size comparable to that of the observed AR. Its corresponding asymmetric BMR (Figure \ref{fig2} (c)) remains similar to the symmetric one. 

In contrast, the complex flux distribution of NOAA 10486 results in its input ($\theta_\pm,\phi_\pm$) that are very close to each other. The resulting polarity separation $\Delta \beta$ is only 0.17$^\circ$, which is much smaller than $\delta_{min}$. This further leads to a significant morphological discrepancy between the realistic AR (Figure \ref{fig2} (d)) and its symmetric BMR approximation (Figure \ref{fig2} (e)), even though key quantities such as total magnetic flux are retained. For the asymmetry case, the small $\Delta \beta$ would produce an abnormal BMR whose evolution significantly differs from that of the realistic NOAA 10486. We therefore adjust the longitudes of the input polarities to correct polarity separation so that $\Delta \beta=\delta_{min}/k$. This correction could reduce the discrepancy, at least in terms of the evolution of the axial dipole strength, and resulting approximated BMR is shown in Figure \ref{fig2} (f).

Since a single abnormal flux source can affect the evolution of an entire solar cycle \citep{Jiang2019, Yeates2023}, the impact of our approximation method on other quantities, such as the polar fields and the equatorial dipole strength, requires further investigation. Improved approximation methods for complex ARs will be explored in future work.

\subsection{Algebraic Quantification of Flux Source Contributions to Axial Dipole Strength}\label{subsec:algebraic}

In addition to the SFT simulation, the algebraic method introduced by \cite{Petrovay2020} provides an efficient way to estimate the contribution of regular BMR to the axial dipole strength at the solar minimum. \cite{Wang2021} further propose a generalized algebraic method that is applicable to ARs or BMRs with arbitrary configurations. This algebraic method directly obtains the axial dipole strength from the magnetic field distribution, thereby avoiding time-consuming SFT simulations. The formulation is
\begin{equation}\label{eq9}
    D = A_0 \int\int B(\lambda,\phi)erf(\lambda /\sqrt{2}\lambda_R)cos\lambda d\lambda d\phi,
\end{equation}
where $D$ is the estimated axial dipole strength at solar minimum for a given magnetic field distribution $B(\lambda,\phi)$, and $\lambda$ denotes latitude. The scaling constant $A_0$ is set to 0.21, following \cite{Wang2021}. The parameter $\lambda_R$, known as the ``dynamo effectivity range", characterizes the combined influence of diffusivity $\eta$ and divergence at the equator $\Delta v$ and is defined as,
\begin{equation}\label{eq10}
    \lambda_{R}=\sqrt{\frac{\eta}{R_{\odot} \Delta v}},
\end{equation}
where $R_{\odot}$ is the solar radius. Equation (\ref{eq9}) has been validated in \cite{Wang2021, Wang2024}. This enables us to apply the method consistently to both realistic ARs and approximated BMRs and to compare the estimated contributions with the observed values for additional verification.

\section{Results} \label{sec:results}
\subsection{Axial Dipole Strength Evolution from Realistic AR Sources}\label{subsec:realsimulation}

\begin{figure}[!ht]
	\centering
	\includegraphics[width=1.0\linewidth]{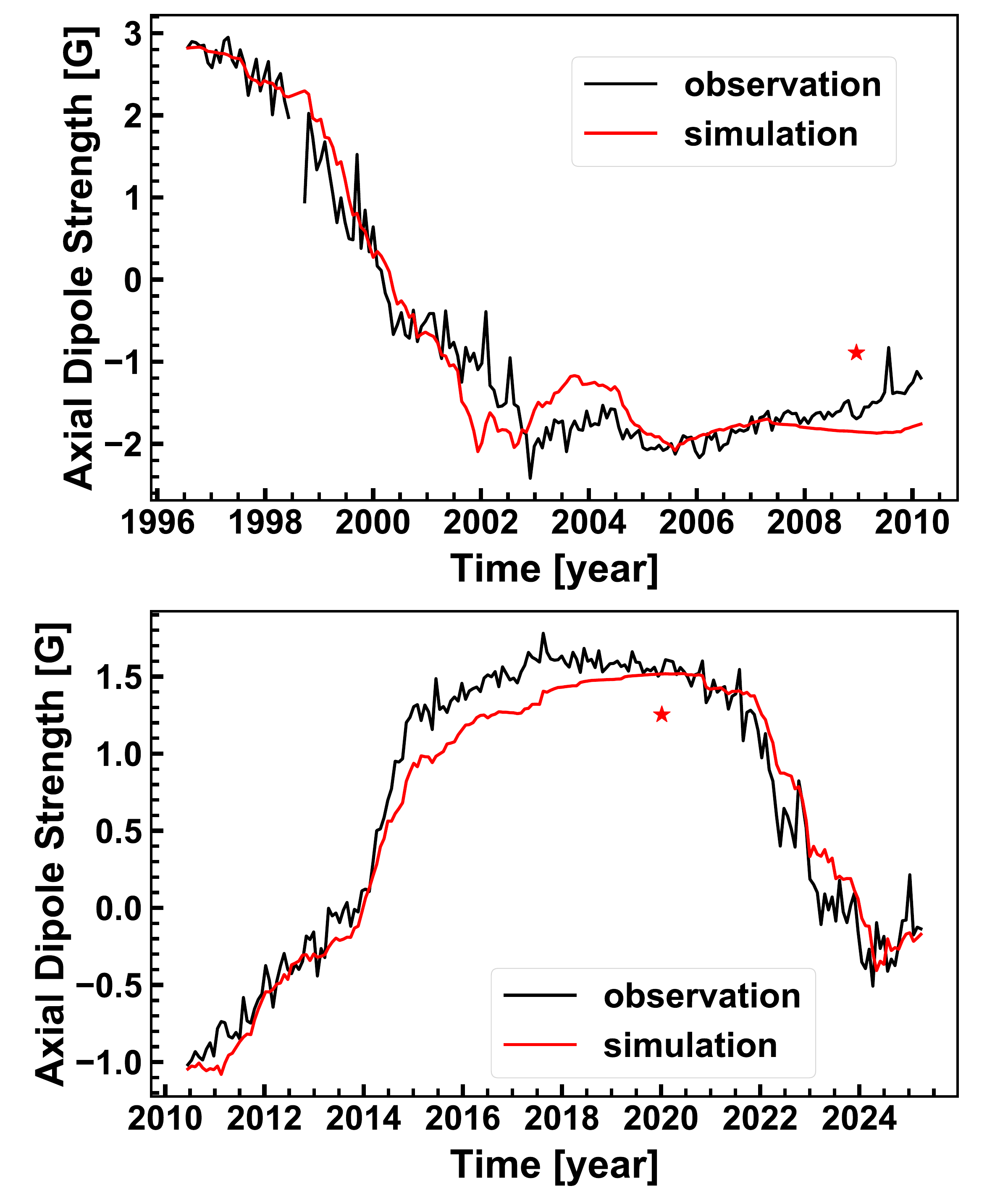}
	\caption{Comparison of the axial dipole strength between observations (black line) and simulations incorporating realistic AR configurations (red line) for the MDI period (upper panel) and HMI period (lower panel). The red stars indicate the axial dipole strength evaluated using the algebraic method (Equation \ref{eq9}) at the ends of solar cycles 23 and 24.}
	\label{fig3}
\end{figure}

Based on the methods in the previous section, we aim to first reproduce the temporal evolution of the axial dipole strength from cycle 23 to the ongoing cycle 25 as the baseline for later comparisons. To avoid the potential artificial influence, simulations are performed separately for the MDI and HMI periods. The CR 1911 (1996 July) and CR 2097 (2010 May) synoptic maps with resolution of $180\times360$ are taken as the initial conditions for the simulations during the MDI and HMI periods, respectively. The transport parameters are set to $\eta=450\ \mathrm{km}^2/\mathrm{s},\ v_0=15\ \mathrm{m/s}\ (\lambda_R=6.02^\circ)$ for the MDI period and $\eta=450\ \mathrm{km}^2/\mathrm{s},\ v_0 = 13\ \mathrm{m/s}\ (\lambda_R=6.46^\circ)$ for the HMI period. Both parameter sets fall within the ranges suggested by \cite{Luo2025}. 

Figure \ref{fig3} compares the observed and simulated axial dipole strengths $D$, computed as
\begin{equation}\label{eq11}
    D=\frac{3}{4\pi}\int_0^{2\pi}\int_0^{\pi}B(\theta,\phi,t)\cos\theta\sin\theta d\theta d\phi, 
\end{equation}
which corresponds to the spherical harmonic coefficient with the spherical harmonic degree $l=1$ and azimuthal order $m=0$. The axial dipole strength is the only model output in this paper. Here, $B(\theta,\phi,t)$ is the observed or simulated surface magnetic field. During the MDI period, the Pearson correlation coefficient is $r=0.97$, and the root-mean-square error (RMSE) is 0.378 G; for the HMI period, $r=0.98$ and RMSE is $0.195$ G. The simulated axial dipole strength variations $\Delta D_{sim}$ are -4.54 G from CR 1912 (1996 August) to CR 2078 (2008 December), and 2.56 G from CR 2097 (2010 May) to CR 2225 (2019 December), compared with the observed variations of -4.416 G and 2.519 G. The corresponding relative errors calculated by ($\Delta D_{sim}-\Delta D_{obs}$)/$\Delta D_{obs}$ are only $2.9\%$ and $1.6\%$. The simulated reversal times also closely match the observations. These results demonstrate that simulations incorporating realistic ARs reliably reproduce the observed axial dipole evolution. 

In particular, this consistency for the HMI period, which spans across two solar cycles, does not rely on incorporating the additional radial diffusion term or varying transport parameters. Even when the same meridional flow speed ($v_0=14\ \mathrm{m/s}$) is used for both MDI and HMI periods, the simulations remain consistent with observations, with only a slight reduction in agreement.


The red stars in Figure \ref{fig3} denote the axial dipole values estimated using the algebraic method (Equation \ref{eq9}). These estimates are obtained by summing the contributions of all identified ARs from CR 1912 to CR 2078 and from CR 2097 to CR 2225, so located at the solar minimums between cycles 23/24 (CR 2078) and 24/25 (CR 2225). Quantitatively, the estimated contributions for the two periods ($\Delta D_{alg}$) are –3.612 G and 2.270 G, respectively. The relative errors with respect to observations, ($\Delta D_{alg}-\Delta D_{obs}$)/$\Delta D_{obs}$, are $18.2\%$ and $9.9\%$, which are acceptable for a rapid and non-simulated evaluation. Even so, the algebraic method provides a sufficiently accurate and efficient means of estimating the axial dipole strength at the solar minimum without running full SFT simulations.

\subsection{Assessing Impacts of Symmetric BMR Approximation} \label{subsec:symmetry}

\begin{figure}[!ht]
	\centering
	\includegraphics[width=1.0\linewidth]{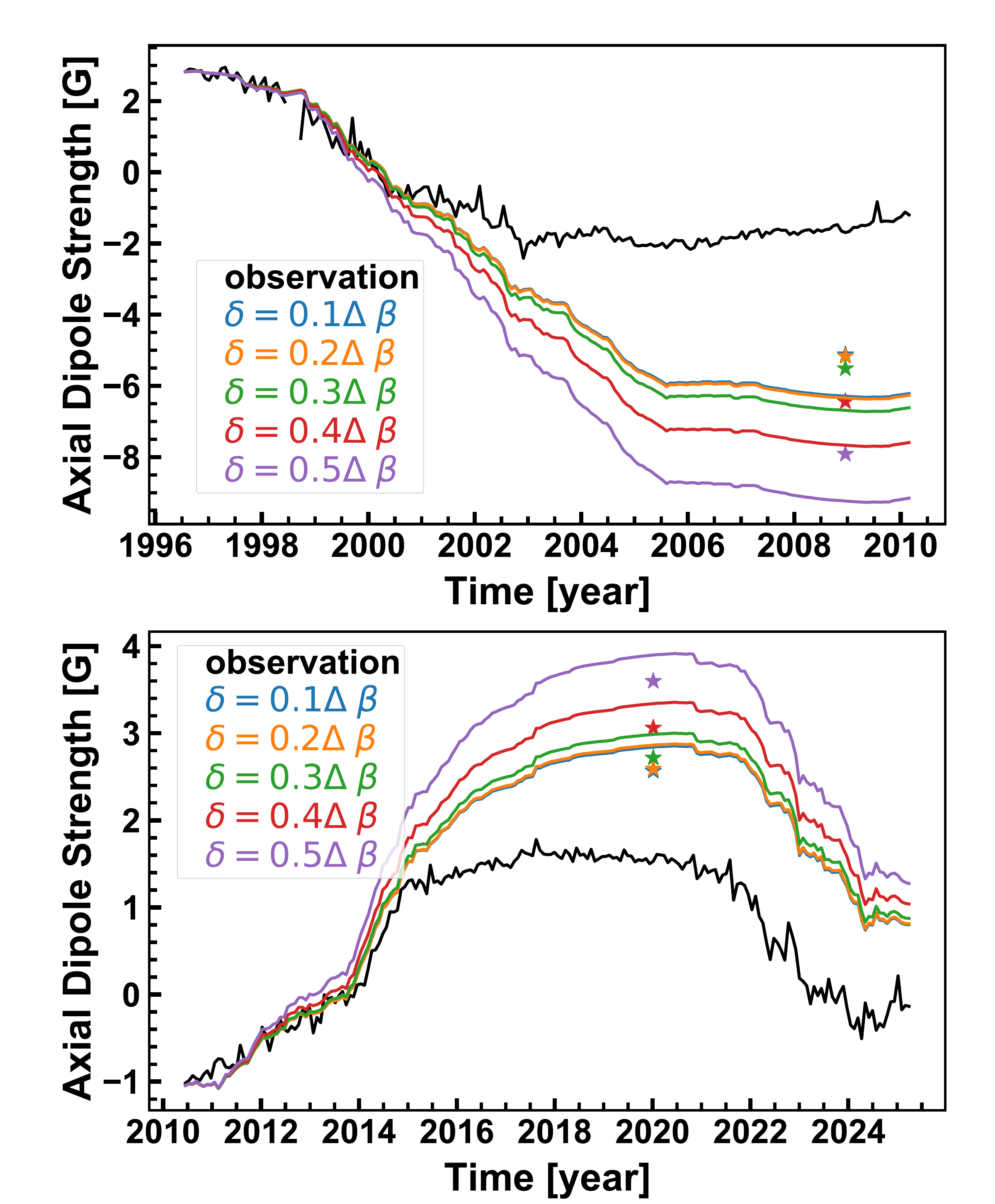}
	\caption{Comparison between observed and simulated axial dipole strength, similar to Figure \ref{fig3} but incorporating symmetric approximated BMR with different proportionality coefficients, $k=$ 0.1, 0.2, 0.3, 0.4, and 0.5.}
	\label{fig4}
\end{figure}

\begin{table*}[!ht]
\centering
\caption{Quantitative Analysis and Comparison of Simulation Results with Incorporating Symmetric BMR and Realistic AR.}
\label{table1}
\begin{tabular}{@{}ccccc@{}}
\toprule
\toprule
& \multicolumn{2}{c}{Relative Error} & \multicolumn{2}{c}{RMSE (G)} \\ \midrule
& MDI period             & HMI period            & MDI period   & HMI period   \\ \midrule
$\delta=0.1\Delta \beta$                 & 129.7\%                & 54.1\%                & 2.918        & 0.984        \\
$\delta=0.2\Delta \beta$                 & 130.9\%                & 55.0\%                & 2.951        & 0.999        \\
$\delta=0.3\Delta \beta$                 & 139.8\%                & 60.0\%                & 3.189        & 1.086        \\
$\delta=0.4\Delta \beta$                 & 164.8\%                & 73.9\%                & 3.864        & 1.335        \\
$\delta=0.5\Delta \beta$                 & 204.6\%                & 96.1\%                & 4.955        & 1.731        \\
realistic AR & 2.9\%                  & 1.6\%                 & 0.378        & 0.195        \\ \bottomrule
\end{tabular}
\end{table*}

In this subsection, we examine how approximating realistic ARs as symmetric BMRs, and varying the size of the approximated BMRs, affect the simulation results. The adopted approximation method is described in Sections \ref{subsubsec:sym} and \ref{subsubsec:ARtoBMR}. The size $\delta$ in Equation (\ref{eq5}) directly determines the size of the approximated BMR and is assumed to scale with the polarity separation $\Delta \beta$, such that a larger proportionality coefficient $k$ corresponds to a larger BMR size with constant unsigned flux. To quantify its influence, we perform simulations with five various coefficients, $k=$ 0.1, 0.2, 0.3, 0.4, and 0.5, while fixing the minimum threshold at $\delta_{min}=2^\circ$. All other simulation parameters follow Section \ref{subsec:realsimulation}. The comparison results are presented in Figure \ref{fig4} and Table \ref{table1}.

As expected, replacing realistic AR configurations with idealized symmetric BMRs substantially overestimates their contributions to the axial dipole strength. This remains true for both direct SFT simulations and algebraic evaluations, and for all values of $k$. The relative overestimation reaches least $129.7\%$ for the MDI period and $54.1\%$ for the HMI period (see Table \ref{table1}), far exceeding the errors obtained when incorporating realistic ARs. Similar overestimates are reported by \cite{Yeates2020, Wang2021, Wang2024}. The smaller overestimate ($24\%$) found by \cite{Yeates2020} for the HMI period may arise from differences in approximation methods and transport parameters.

A clear trend is that decreasing the proportionality coefficient, i.e., reducing the angular width $\delta$ of approximated BMR, monotonically reduces the overestimation. Because $\delta_{min}=2^{\circ}$ and most ARs have $\Delta \beta < 10^\circ$, the improvement by reducing the coefficient from 0.2 to 0.1 is limited, as shown by the almost overlapping yellow and blue curves in Figure \ref{fig4}. Even so, the relative error and the RMSE remain significantly higher than in simulations using realistic ARs. 

The suppression of the overestimation with decreasing $\delta$ indicates that more spatially concentrated flux sources produce smaller contributions to the axial dipole strength, and thus to the polar fields. In other words, simultaneously reducing only the size of two BMR polarities increases the amount of magnetic flux crossing the equator. Extrapolating this trend suggests that a point bipole, or doublet, used by \cite{Wang1989, Sheeley1985, DeVore1987b} would be a more idealized source term. However, such a Dirac-delta-like source cannot be implemented in our SFT code because of spatial resolution limits and the spherical harmonic decomposition algorithm. 

A similar reduction in the axial dipole strength contribution is produced by inflows toward ARs, as included in the models of \cite{Jiang2010} and \cite{Martin2017}. By suppressing flux dispersal, the inflow effectively inhibits AR flux crossing the equator, consequently weakening the axial dipole strength. Alternatively, \cite{Cameron2010, Jiang2020} attribute the good performance of their model, achieved by multiplying the tilt angle by a factor of 0.7, to the potential effect of the inflow.

It is also worth noting that some previous studies \citep{Cameron2010, Jiang2020} incorporate BMRs with a constant $\delta$ and rescale $B_0$. Large BMRs are therefore compressed into overly compact configurations, while small ones become excessively diffuse, under the condition of constant unsigned magnetic flux. Although this approach also avoids issues with small BMRs that cannot be resolved due to spatial resolution limits, our results suggest that it overestimates the contribution of small BMRs and underestimates that of large ones. The implications of this BMR approximation method need further investigation.

\subsection{Assessing Impacts of Asymmetric BMR Approximation}\label{subsec:asymmetry}

\begin{figure}[!ht]
	\centering
	\includegraphics[width=1.0\linewidth]{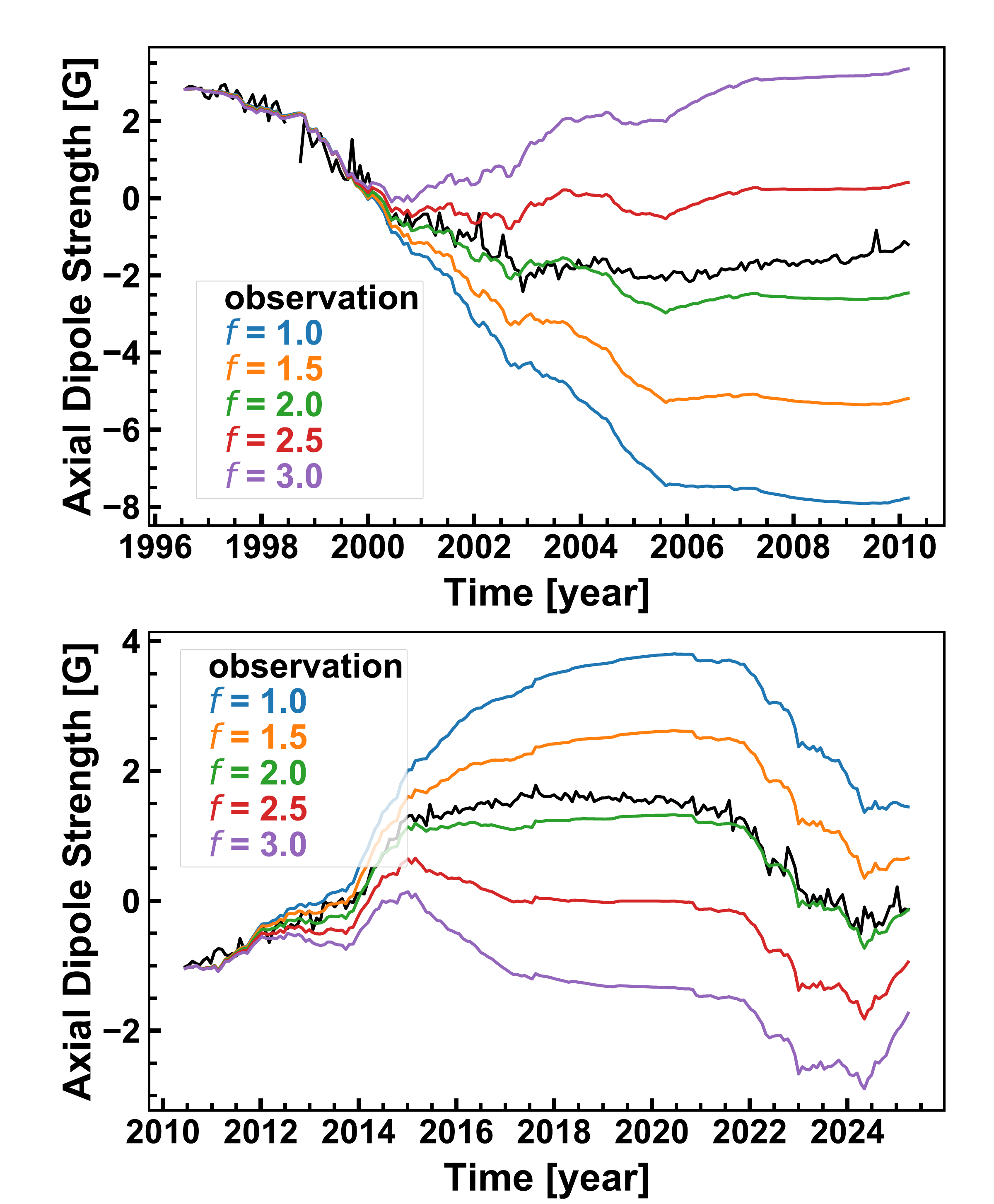}
	\caption{Similar to Figure \ref{fig3}, but incorporating asymmetric approximated BMR with $k=0.4$ and different asymmetry factors, $f=$ 1.0, 1.5, 2.0, 2.5, and 3.0.}
	\label{fig5}
\end{figure}

\begin{figure}[!ht]
	\centering
	\includegraphics[width=1.0\linewidth]{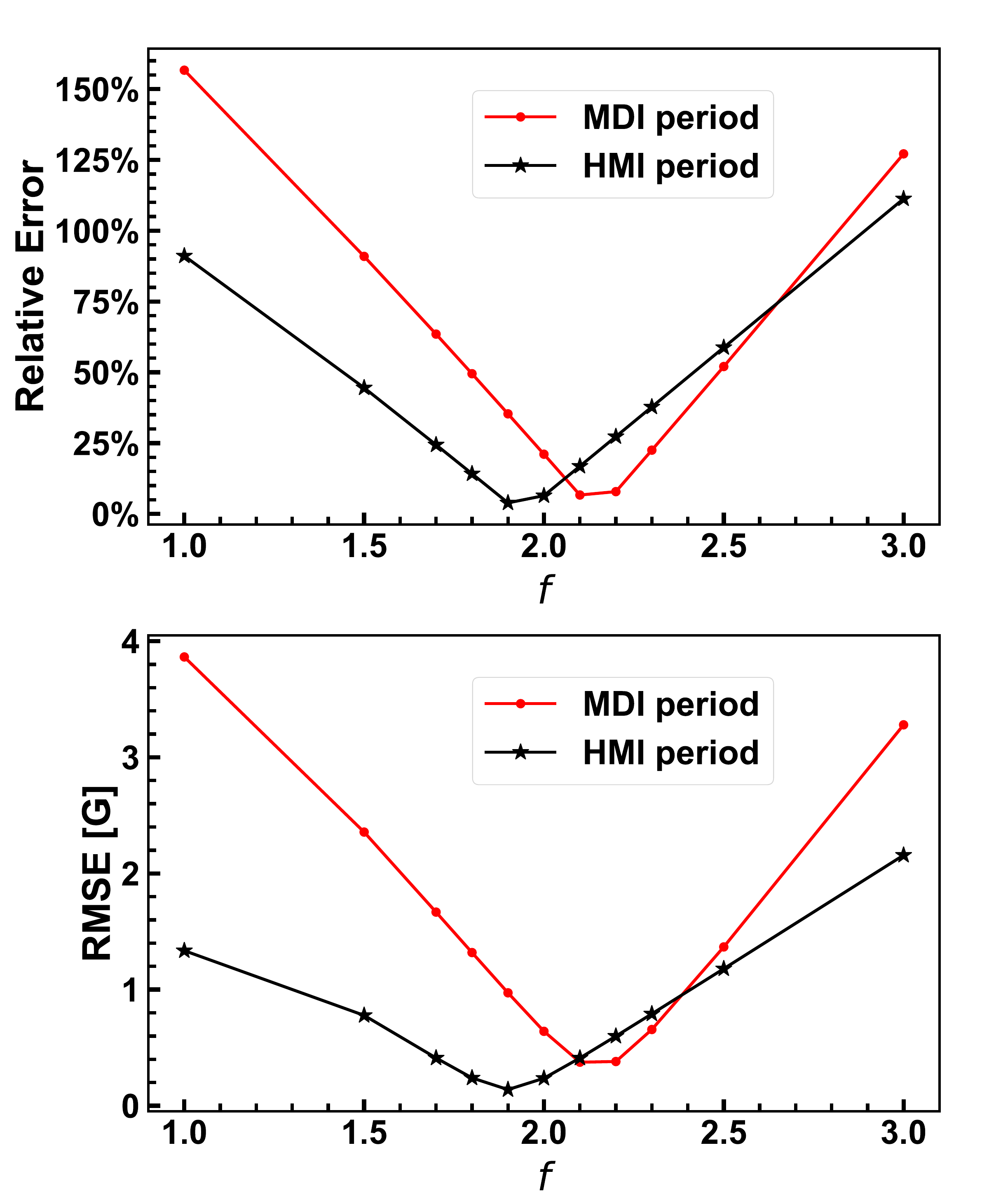}
	\caption{Variation of quantitative indices with $f$ values. The upper and lower panels correspond to the relative error and root-mean-square error (RMSE) of the simulated axial dipole strength, respectively. The red lines with circle markers represent the results for the MDI period, while the black lines with star markers represent the results for the HMI period.}
	\label{fig6}
\end{figure}

To assess the influence of asymmetric BMR approximation, we directly adjust the size ratio $f$ to generate BMRs with different asymmetry degrees, unlike \cite{Iijima2019}, who realizes this indirectly through the observed morphological sunspot area asymmetry \citep{Tlatov2014}. All other parameters, including the total unsigned magnetic flux, are kept as close as possible to those of the original ARs, allowing the effects of asymmetry to be isolated. Here, we use the same transport parameter setup as in \ref{subsec:realsimulation}. We allow the free parameter $f$ to vary from 1.0 (symmetry) to 3.0 (strong asymmetry). Results for the commonly used coefficient $k=0.4$ \citep{vanBallegooijen1998, Iijima2019} are shown as representative examples.

Figure \ref{fig5} presents the simulated axial dipole strength for five representative values of $f$. Compared with the symmetric BMR case ($k=0.4,\ f=1$), introducing morphological asymmetry would significantly suppress the overestimation of the axial dipole strength at the solar cycle minimum. Moreover, the axial dipole strength decreases nearly linearly with increasing $f$. Similar behavior is found for the cases with other values of $k$.

This linear reduction of axial dipole strength with the increase in $f$ can be understood by comparing different ways of modifying the source configuration. In our approach, morphological asymmetry is introduced by enlarging the size of the following polarity while keeping the leading polarity unchanged. This behavior indicates that enlarging only one polarity reduces its effective contribution to the polar field. This applies equally to the leading polarity. In other words, increasing the size of only one polarity tends to reduce the amount of its flux ultimately transported to the polar regions, likely through enhanced magnetic flux crossing the equator. 

By contrast, enlarging both polarities simultaneously (the case in Section \ref{subsec:symmetry}) increases the axial dipole strength. This implies that the influence of the leading polarity at lower latitudes exceeds that of the following polarity at higher latitudes, since the leading polarity typically contributes negatively to the axial dipole strength. This interpretation is consistent with \cite{Petrovay2020}. Together, these results highlight the sensitivity of SFT simulations to the size of the approximated BMRs.

Excessive asymmetry, however, leads to unphysical behavior. For instance, the simulated axial dipole strength fails to reverse eventually within a single solar cycle during both the MDI and HMI periods with $f=3.0$, as shown in Figure \ref{fig5}. The evolution deviates strongly near solar maximum and approaches the initial value again at solar minimum. A similar phenomenon is reported by \cite{Iijima2019}, who find that introducing strong asymmetry would yield opposite contributions. These results indicate that although incorporating asymmetry is important for SFT simulations, it must be constrained.

We perform a systematic parameter sweep over the range of $f=1.0$ to 3.0 to determine the value that best matches the observations. The relative errors and RMSE between the observed and simulated axial dipole strength are used as quantitative evaluation metrics. The values of these metrics for different values of $f$ and fixed $k=0.4$ are displayed in Figure \ref{fig6}. Both quantitative metrics decrease initially and then rise as asymmetry becomes stronger. For both the MDI and HMI periods, the minima occur near $f\approx2$ and are close to the values obtained when incorporating realistic AR configurations, with only minor differences between the two periods. This trend is also evident in Figure \ref{fig5}, where the simulation with $f=2$ reproduces the axial dipole evolution of the realistic AR case, and thus the observations. 

Although the optimal value of $f$ depends on the choice of $k$, our goal here is to provide a practical parameter choice that enables approximated BMRs to reproduce results comparable to those obtained with realistic ARs. In this sense, the combination $f\approx2$ and $k=0.4$, represents a recommended and effective choice, at least for solar cycles 23 to 25. 

\section{Conclusion and Discussion} \label{sec:conclusion}

In this study, we present a systematic quantitative assessment of how magnetic flux source configuration affects SFT simulations. Using realistic AR configurations, we first demonstrated that our SFT model successfully reproduces the observed evolution of the axial dipole strength over multiple solar cycles. These simulations also confirm the validity of the algebraic method of \cite{Wang2021} as an efficient tool for estimating the axial dipole strength at the solar minimum without numerical integration.

Compared with incorporating realistic ARs, replacing them with idealized symmetric BMRs leads to a substantial overestimation of their contribution to the axial dipole strength during solar minima, consistent with earlier findings \citep{Yeates2020, Wang2021}. This overestimation could partially be suppressed by decreasing the size of the approximated BMR, but cannot be eliminated. In contrast, introducing morphological asymmetry could significantly reduce overestimation of the simulated axial dipole strength. This improvement strongly depends on the asymmetry factor $f$. We find that adopting $f\approx2$ with $k=0.4$ yields axial dipole strength that closely match those obtained using realistic ARs. This provides a practical constraint for improving SFT simulations that rely on approximated BMRs.

Our results complement and refine earlier work about asymmetric BMRs. \cite{Iijima2019} indirectly adjust $f$ through its correlation with the observed sunspot area asymmetry of 0.4 from \cite{Tlatov2014}, obtaining an optimal value of $f=2.28$. However, their indirect optimization and relative large parameter step cause discontinuities in the inferred asymmetry range, leading to a jump from $f=2.28$ to $f=1.54$. Our direct and systematic exploration of $f$ bridges this gap and yields a similar but more robust estimate. We also find that optimal $f$ is slightly different for MDI and HMI periods. This is consistent with the indirect observation that sunspot area asymmetry varies from cycle to cycle \citep{Tlatov2014, Tlatov2015}. A remaining limitation is the lack of direct observational constraints on AR size asymmetry itself. With high-resolution AR databases now available, systematically extracting $f$ from magnetic observations would provide a valuable constraint and an important direction for future work.

The overestimation associated with incorporating symmetric BMRs is slightly larger during the MDI period than during the HMI period. Because the ARISE database has been calibrated to minimize the instrumental inconsistencies between MDI and HMI, the artificial effect from instruments could be largely excluded. We are therefore more inclined to attribute this discrepancy primarily to the intrinsic difference between various solar cycles. In particular, \cite{Jiang2015} report that a number of ARs with abnormal tilt angles occurred during cycle 23 (the MDI period), which might contribute to the stronger overestimation during this period. In addition, other parameters not included in our simulations, such as the temporal variation of inflow and meridional flow \citep{Zhao2004, Hathaway2011}, might also contribute to the cycle-to-cycle differences.

The axial dipole strength is only one aspect of the solar surface magnetic field. Although the optimization based on the axial dipole strength is sufficient for solar cycle prediction and related applications, it does not ensure accurate reconstruction of other important features, such as polar-field strength, latitudinal distribution of surface magnetic fields, or equatorial dipole evolution. The parameter constrained in this paper, therefore, provides a foundation for extending the analysis to additional solar surface field investigations, including higher-order multipoles and lower spatial-scale structures like \cite{Luo2023, Luo2024}. Such extensions will be essential for achieving a more comprehensive and physically consistent characterization of surface flux transport.

\begin{acknowledgments}
We thank the anonymous referee for the valuable comments and suggestions, which improved the overall quality of this paper. The research is supported by the National Natural Science Foundation of China (grant Nos. 12425305, 12350004, and 12173005) and China's Space Origins Exploration Program. The SDO/HMI data are courtesy of NASA and the SDO/HMI team. SOHO is a project of international cooperation between ESA and NASA.
\end{acknowledgments}






\bibliography{power_spectra}{}
\bibliographystyle{aasjournalv7}



\end{document}